\documentclass[%
reprint,
superscriptaddress,
amsmath,amssymb,
aps,
floatfix,
]{revtex4-2}

\usepackage[
activate={true,compatibility},
final,
tracking=true,
kerning=true,
nopatch=footnote]{microtype}
\usepackage{mathtools}
\usepackage{amsfonts}
\usepackage{nicefrac}
\usepackage{physics}
\usepackage{color}
\usepackage{graphicx}
\usepackage{dcolumn}
\usepackage{booktabs}
\usepackage{tabularx}
\usepackage{multirow}
\usepackage{xcolor}
\usepackage[normalem]{ulem} 
\usepackage[noend]{algpseudocode}
\usepackage{amsthm}

\DeclareRobustCommand{\rchi}{{\mathpalette\irchi\relax}}
\newcommand{\irchi}[2]{\raisebox{\depth}{$#1\chi$}} 

\makeatletter
\renewcommand\@makecaption[2]{%
  \par
  \vskip\abovecaptionskip
  \begingroup
   \small\rmfamily
    \begingroup
     \samepage
     \flushing
     \let\footnote\@footnotemark@gobble
     \@make@capt@title{#1}{#2}\par
    \endgroup
  \endgroup
  \vskip\belowcaptionskip
}
\makeatother

\newcolumntype{M}{>{\centering}X}
\newcolumntype{Y}{>{\hsize=.115\textwidth\arraybackslash}X}
\newcolumntype{C}{>{\hsize=.115\textwidth\centering\arraybackslash}X}
\newcolumntype{R}{>{\hsize=.115\textwidth\raggedleft\arraybackslash}X}

\newtheorem{theorem}{Observation}

\begin{document}


\title{Complete quantum-inspired framework for computational fluid dynamics}


\author{Raghavendra D. \surname{Peddinti}}
\affiliation{Quantum Research Center, Technology Innovation Institute, Abu Dhabi, UAE}
\affiliation{Department of Mathematics, ETH Zurich, 8092 Zurich, Switzerland}
\author{Stefano \surname{Pisoni}}
\affiliation{Quantum Research Center, Technology Innovation Institute, Abu Dhabi, UAE}
\affiliation{Institute for Quantum-Inspired and Quantum Optimization, Hamburg University of Technology, Germany}
\author{Alessandro \surname{Marini}}
\affiliation{Propulsion and Space Research Center, Technology Innovation Institute, Abu Dhabi, UAE}
\author{Philippe \surname{Lott}}
\affiliation{Propulsion and Space Research Center, Technology Innovation Institute, Abu Dhabi, UAE}
\author{Henrique \surname{Argentieri}}
\affiliation{Propulsion and Space Research Center, Technology Innovation Institute, Abu Dhabi, UAE}
\author{Egor \surname{Tiunov}}
\affiliation{Quantum Research Center, Technology Innovation Institute, Abu Dhabi, UAE}
\author{Leandro \surname{Aolita}}
\affiliation{Quantum Research Center, Technology Innovation Institute, Abu Dhabi, UAE}
\affiliation{Federal University of Rio de Janeiro, Caixa Postal 652, Rio de Janeiro, RJ 21941-972, Brazil}

\date{\today}

\begin{abstract}
Computational fluid dynamics is both a thriving research field and a key tool for advanced industry applications. The central challenge is to simulate turbulent flows in complex geometries, a compute-power intensive task due to the large vector dimensions required by discretized meshes. We present a full-stack method to solve for incompressible fluids with memory and runtime scaling poly-logarithmically in the mesh size. Our framework is based on matrix-product states, a powerful compressed representation of quantum states. It is complete in that it solves for flows around immersed objects of arbitrary geometries, with non-trivial boundary conditions, and self-consistent in that it can retrieve the solution directly from the compressed encoding, i.e. without ever passing through the expensive dense-vector representation. This machinery lays the foundations for a new generation of potentially radically more efficient solvers of real-life fluid problems.
\end{abstract}

\maketitle


\section{Introduction} \label{section:Introduction} 
The Navier-Stokes equations remain one of the biggest open problems in physics~\cite{fefferman2000existence}, with closed-form solutions known only in restricted cases.
This has given rise to the field of Computational Fluid Dynamics (CFD)
~\cite{DNS_orszag1972numerical}.
Most 
CFD methods rely on a problem-specific discretization of the spatial domain, a {\it mesh}. 
There, the velocity and pressure fields are represented by vectors whose dimension is given by the mesh size.
However, the precision required to capture relevant dynamics often translates into prohibitively high dimensions. 
For instance, for an accurate description of turbulent flows, the ratio between the largest and the smallest length scale in the mesh must grow with the Reynolds number~\cite{kolmogorov1941local,pope_2000}. The resulting mesh can then have sizes that render the problem intractable for standard methods.
This is a manifestation in CFD of the infamous \textit{curse of dimensionality}, 
the most serious limitation of state-of-the-art 
mesh-based solvers.

A similar limitation arises in simulations of quantum systems, described by vector 
spaces exponentially large in the number $N$ of particles \cite{eisert2010colloquium, poulin2011quantum}.
There, sophisticated tensor-network techniques~\cite{schollwock2011density,orus2014practical,verstraete2008matrix} have been developed to tackle the problem under the assumption of low entanglement, i.e. non-factorability  of the quantum state 
~\cite{horodecki2009quantum}.
The best-known example is matrix-product states (MPSs)~\cite{White_PRL_1992,Vidal_PRL_2003}, also known as tensor trains~\cite{oseledets2011tensortrain}.
These represent 
state vectors as a 1D array of $N$ matrices whose maximal size (called {\it bond dimension}) depends on the amount of entanglement.
Hence, low-entangled states of 1D systems can be exponentially compressed by MPSs~\cite{eisert2010colloquium,schollwock2011density,orus2014practical,verstraete2008matrix}. 
In fact, nowadays, MPSs and their tensor-network extensions provide  the most powerful framework for simulating complex quantum dynamics~\cite{What_limits_the_sim_of_QC_Stoudenmire,Stoudenmire_IBM_simulation}.  


In view of this, MPSs  have been applied to a variety of other high-dimensional problems~\cite{Image_Compression,probabilistic_modeling,Tomography_with_TN_Aolita,kurmapu2022reconstructing,option_pricing,Greens_function_MPS,truong2023tensor}.
In particular, turbulent fluids in simple geometries have been observed to also admit an efficient MPS description~\cite{gourianov2022quantum,kiffner_jaksch2023tensor,kornev2023chemicalmixer,gourianov2022_thesis}, remarkably. 
The rationale is that the energy cascade mechanism~\cite{kolmogorov1941local}, whereby energy transfer takes place only between adjacent spatial scales, may play a similar role to local interactions in 1D quantum systems.
This may open a new arena for CFD solvers. 
Nevertheless, for this to happen, compression efficiency should be combined with versatility to describe immersed objects of diverse geometries, with non-trivial boundary conditions. 
Moreover, a practical solver should also allow for accessing the solution directly in the MPS encoding, without passing through the dense vector.
Otherwise, potential speed-ups in solving the problem may be lost at evaluating the solution. 
It is an open question whether these requirements can be harmoniously met and, most importantly, if there are settings of practical relevance where this can be achieved with a moderate bond dimension. 

Here, we answer these questions in the affirmative.
We deliver a complete MPS-based toolbox for simulating incompressible fluids, with complexity logarithmic in the mesh size and polynomial in the bond dimension.
This is achieved via three major innovations: First, 
{\it approximate MPS masks} 
to encode object shapes into MPSs of low bond dimensions.
Second, 
a native time-integration scheme incorporating the masks as built-in features. This is based on finite-differences as in~\cite{gourianov2022quantum,kiffner_jaksch2023tensor}, but it operates with open boundary conditions, as in~\cite{kornev2023chemicalmixer}, crucial to treat immersed objects without too large spatial domains. 
Interestingly, our integrator embeds open boundary conditions into a matrix-product differential operator
.
 
\begin{figure*}[t]
    \includegraphics[width=\textwidth]{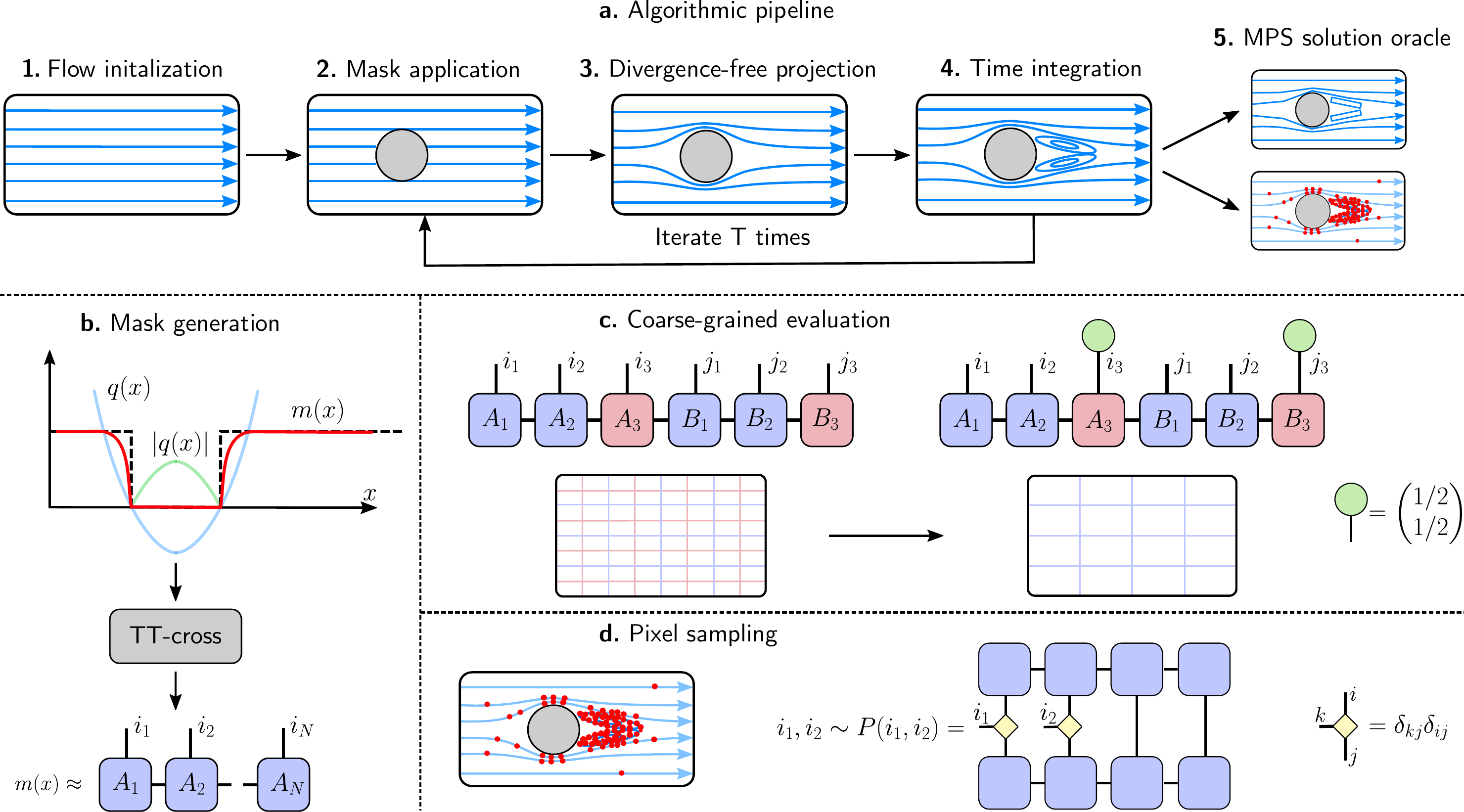}
     \caption{{\bf Schematics of the framework.} {\sf \textbf{a}}. Main steps of the solver, featuring the paradigmatic steady flow around a cylinder. The first step is the flow initialization with constant velocity. The second one is the MPS mask application, which nullifies the velocity field inside the object
     . This produces an artificial field that is not divergence-free. To correct this, in the third step, we project the field back onto the divergence-free manifold
     .
     The fourth step is the time integration over a small evolution time, using an explicit Euler method. The last three steps are then iterated $T$ times to get the final MPS field at time $T$
     .
     Finally, the fifth step is the MPS solution oracle. The two boxes there represent the main operation modes: coarse-grained evaluation and pixel sampling.
     {\sf \textbf{b}}. Construction of approximate MPS masks (for a 1D object for clarity).
     Given a smooth function $q(x)$ (in light blue, its absolute value in green) that vanishes on the boundary of the object, we define $m(x)=1-e^{-\alpha\left(q(x)+|q(x)|\right)}$ (red solid). This approximates the object's indicator function $\theta(x)$ (black dashed) up to a tunable small error (through $\alpha$). Next, we apply the TT-cross algorithm to produce an MPS approximation to $m(x)$ from few queries of it. 
     {\sf \textbf{c}}. Coarse-grained evaluation: a 4-bit MPS, encoding an average over the smallest (pink) spatial scale
     , is obtained directly from a 6-bit MPS through simple contractions with constant single-site gadget tensors in green. 
     {\sf \textbf{d}}. Pixel sampling:  mesh cells (red points) are randomly chosen with probabilities proportional to the squared vorticity field at those points. The key tool are the marginals of the joint probability distributions over the binary representation of the cell positions. These allow one to sample the entire string bit by bit (see main text) and can be efficiently obtained directly from the MPS. In the figure, a 2-bit marginal is obtained through simple single-site contractions of two copies of 4-bit MPS and the gadget tensors in yellow. Remarkably, neither operation mode of the oracle requires field evaluations in the dense-vector representation (see main text).}
    \label{fig:outline}
\end{figure*}

Third, 
an {\it MPS solution oracle} for querying the solution directly from the MPS, bypassing dense-vector evaluations.  
This features two main operation modes: \textit{coarse-grained evaluation} and \textit{pixel sampling}. 
In particular, the latter generates random points weighted by the squared magnitude of a target property (vorticity, pressure, etc.), mimicking the measurement statistics on a quantum state
. This is ideal for Monte-Carlo simulations.

We validate the framework by reproducing text-book regimes of flows around a cylinder and by comparing our solutions for flows around a square, and even a collection of squares, with those of the popular commercial solver \texttt{Ansys Fluent}. 
We observe a remarkable agreement with the latter.
To our knowledge, this is the first MPS algorithm for differential-equation systems validated to such extents.
Finally, we venture into flows around a real-life airfoil from the NACA data set. MPSs of 18800 parameters are enough for an accurate simulation with a mesh of $2^{19}=524288$ pixels, an impressive 27.9X compression. Among others, we use pixel sampling to randomly explore the regions of highest vorticity around the wing, without evaluating a single component of the $2^{19}$-dimensional vorticity vector field itself.
These tools may be relevant for aero- and hydro-dynamic design optimizations.

\section{Preliminaries}
\subsection{Incompressible Navier-Stokes equations} \label{subsection:NS equations}
We solve the incompressible Navier-Stokes equations in the absence of external forces (for simplicity):
\begin{subequations}\label{eq:NS}
\begin{align} 
        \frac{\partial{\boldsymbol{v}}}{\partial{t}} + (\boldsymbol{v} \cdot \nabla )\, \boldsymbol{v} &= -\frac{1}{\varrho} \nabla p + \nu\, \nabla^2\, \boldsymbol{v}, \label{eq:NS_eq_a} \\
        \nabla \cdot \boldsymbol{v} &= 0, \label{eq:NS_eq_b}
\end{align}
\end{subequations}
where $\boldsymbol{v}=\boldsymbol{v}(\boldsymbol{x},t)$ and $p=p(\boldsymbol{x},t)$ are respectively the velocity and pressure fields at position $\boldsymbol{x}$ and time $t$, $\varrho$ is the  density, and $\nu$ the kinematic viscosity. Eqs.~\eqref{eq:NS_eq_a} and~\eqref{eq:NS_eq_b} follow respectively from   momentum and mass conservations~\cite{monin2007,monin2013statistical,pope_2000}. They give rise to a variety of behaviours, ranging from laminar to turbulent flows. 
The flow regime is determined by the Reynolds number, Re $= \frac{U L}{\nu}$, where $U$ and $L$ are characteristic velocity and length scales of the problem~\cite{pope_2000}.
This dimensionless parameter quantifies the ratio between the inertial [$(\boldsymbol{v}\cdot\nabla)\,\boldsymbol{v}$] and dissipative [$\nu\,\nabla^2\boldsymbol{v}$] force terms.
At high Re, the nonlinear inertial term dominates and the flow is highly chaotic and turbulent. Instead, at low Re, the dissipative term dominates and the flow is stable and laminar.

\subsection{Field encoding with matrix product states} \label{subsection:MPS}
 We encode scalar functions -- such as the velocity components $v_x(\boldsymbol{x})$ and $v_y(\boldsymbol{x})$, or the pressure $p(\boldsymbol{x})$ -- into matrix-product states (MPSs)~\cite{schollwock2011density,orus2014practical,verstraete2008matrix}. 
We discretize the 2D domain into a mesh of $2^N$ points, specified by $N=N_x+N_y$ bits.
There, we represent a continuous function $v(x,y)$ by a vector of elements $v_{\boldsymbol i,\boldsymbol j} := v(x_{\boldsymbol i},y_{\boldsymbol j})$, where the strings ${\boldsymbol i}:=(i_1,i_2, \hdots i_{Nx})$ and ${\boldsymbol j}:=(j_1,j_2, \hdots j_{Ny})$ give the binary representation of $x_{\boldsymbol i}$ and $y_{\boldsymbol j}$, respectively.
Then, we write each $v_{\boldsymbol i ,\boldsymbol j}$ as a product of $N$ matrices:
\begin{equation}
\label{eq:MPS}
    v_{\boldsymbol i,\boldsymbol j} = A_1^{(i_1)} A_2^{(i_2)}\dots A_{N_x}^{(i_{N_x})}\,B_1^{(j_1)} B_2^{(j_2)}\dots B_{N_y}^{(j_{N_y})}.
\end{equation}
This is the MPS form (see also Fig.~\ref{fig:outline}c). The indices $i_k$ and $j_k$ are referred to as \textit{physical indices} of the matrices $A_j$ and $B_j$, respectively. 
Note that, the first $N_x$ matrices correspond to $x_{\boldsymbol i}$ and the remaining $N_y$ ones to $y_{\boldsymbol j}$, as in~\cite{Greens_function_MPS,kiffner_jaksch2023tensor}. 
However, other arrangements are possible~\cite{Erika_vlasov_eq,Image_Compression,gourianov2022quantum}.
The bond dimension $\rchi$ is defined as the maximum dimension over all $2\,N$ matrices used.  
Importantly, the total number of parameters is at most $2\,N\,\rchi^2$. Hence, if $\rchi$ constant, the MPS provides an exponentially compressed representation of the $2^N$-dimensional vector. Moreover, instead of fixing all virtual dimensions at $\rchi$, we dynamically adapt each site's dimension to the amount of inter-scale correlations. This typically reduces the number of parameters well below the bound $2\,N\,\rchi^2$.

\section{Complete MPS framework for CFD} \label{section:Description of the framework}
Our toolbox is summarized in Fig.~\ref{fig:outline}.
The details of the Navier-Stokes equations as well as of the MPS encoding of velocity and pressure fields are given in the Methods. 
Our presentation is restricted to the 2D case for simplicity, but each and all of the components of the framework can be straightforwardly generalized to the 3D case.

\subsection{Approximate MPS masks} \label{subsection: Mask generation}
If a rigid body is immersed in a fluid at a fixed position, the velocity field must be null both inside the object and on its boundary (no-slip condition).
This is a crucial requirement that, to our knowledge, has not been  addressed in the literature of quantum-inspired solvers. 
To incorporate it, we introduce the notion of {\it approximate MPS masks}  (see Fig.~\ref{fig:outline}b).  These are MPSs encoding functions that approximate the target indicator function $\theta$ of the object -- i.e. the function equal to zero within the object and to one outside it -- up to arbitrary, tunable precision. We apply these masks on the MPS representing the field by simple element-wise multiplication.

To build such MPSs, we first consider a scalar function $q(\boldsymbol{x})$, with $\boldsymbol{x}\in \mathcal{R}^D$, such that it $i$) vanishes on the boundary of the $D$-dimensional object in question, $ii$) takes negative values in the interior of the object, and $iii$) tends to infinity far away from the boundary. For example, for a circular object of unit radius in $D=2$, $q(x,y) = x^2 + y^2 - 1$ does the job. Using such boundary functions, we define the smooth function
\begin{equation} \label{eq:mask}
    m(\boldsymbol{x}) \coloneqq 1 - e^{- \alpha \left(q(\boldsymbol{x})
            + | q(\boldsymbol{x}) |\right)},
\end{equation}
with $\alpha > 0$. By construction, $m$ approximates $\theta$ increasingly better 
for growing values of $\alpha$.
In particular, the $l_2$ distance between $m$ and $\theta$ decreases with $\alpha$ as $1/\sqrt{\alpha}$ (see App.~\ref{app:mask}).
The next step is to obtain an (approximate) MPS representation of $m$. For elementary functions there are analytical constructions for that~\cite{Oseledets_2012_real_functions,garcia_ripoll2021quantum}. However, in general, one must resort to numerical techniques. 

\begin{figure}[t!]
\centering
    \includegraphics[width=1\columnwidth]{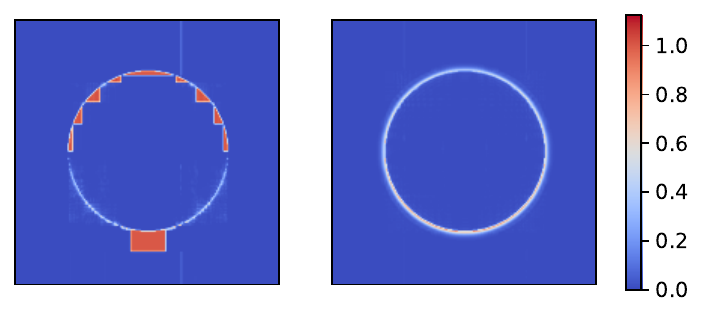}
    \caption{{\bf Mask errors for the cylinder}. The left and right panels show the differences (in absolute value) between the exact indicator function $\theta$ and the MPS masks obtained from the TT-cross approximation directly to $\theta$ and to its smooth approximation $m$, respectively. The domain discretization is $N_x=N_y=11$ bits both horizontally and vertically (we show only a 6X close-up on the object); and the bond dimension is $\rchi = 30$. Clearly, the deviations seen in the left-hand panel are much larger, not only close to the object's boundary but also far from it (note the thin vertical lines on the right-hand side of the panel).  
    That is, the intermediate step through $m$ leads to masks of significantly higher accuracy. This is crucial for the solver, which applies the mask at every time iteration.}
    \label{fig:Mask_22bits}
\end{figure}

The specific technique we use is the standard TT-cross approximation~\cite{OSELEDETS201070TTcross}. 
This generates an MPS approximation of $m$, up to controllable $l_2$ distance $\epsilon$, from few evaluations of $m$. In App.~\ref{app:mask}, we  observe that $\epsilon = 10^{-3}$ is enough for high-quality masks for the cylinder in meshes of up to $2^{30}$ cells. The 30-bit MPS masks generated there have $\chi=30$ and a total of 24200 parameters, corresponding to a compression factor of more than 44360, remarkably.  
Our observations consistently show that the TT-cross approximation directly on the discontinuous function $\theta$ produces significantly worse masks than via the intermediate approximation $m$. This is shown in full-resolution in Fig.~\ref{fig:Mask_22bits} for a 22-bit mesh. (The 30-bit case is to large for full resolution evaluation, see App.~\ref{app:mask}.)

\subsection{Time evolution} \label{subsection: Numerical solver}
We discretize the differential operators in Eqs.~\eqref{eq:NS} using finite differences with an 8-th order stencil~\cite{Fornberg1988GenerationOF_8th_order_stencil}. We use open boundary conditions (BCs), necessary to treat immersed objects using limited spatial domains. 
More precisely, we apply inlet-outlet BCs~\cite{weinan1995projection} at the left- and right-hand edges, where the fluid enters and exits the domain, respectively; and use closed BCs at the upper and lower edges.
We note that, in Ref.~\cite{kornev2023chemicalmixer}, inlet-outlet BCs are enforced via application of a mask. 
Instead, we embed them directly into the differential operator. 
This operator can in turn be analytically recast as a matrix product operator (MPO), as shown in Refs. ~\cite{garcia_ripoll2021quantum,Oseledets_Operators_2010,kazeev2012low} for simpler BCs. As a result, we obtain differential MPOs with inlet-outlet BCs as built-in features, with low bond dimensions, remarkably (see App.~\ref{app:time evolution}). With the differential MPOs in place, we can perform the time evolution.

The first step (see Fig.~\ref{fig:outline}a) is to initialize the velocity MPSs [see Eq.~\eqref{eq:MPS}] as a constant vector field (flow far away from the object). 
Next, we loop $T$ times over steps 2 to 4, with $T$ the desired evolution time.
Step 2 is the object mask application, described above. 
Steps 3 and 4 consist respectively of projecting the resulting MPS velocity field onto the divergence-free manifold, defined by Eq.~\eqref{eq:NS_eq_b}, and integrating over a small time step $\Delta\,t$. We implement these two steps with Chorin's projection method~\cite{chorin1967numerical,chorin1968numerical}.
That is, we first solve for the pressure from a Poisson equation, and use that pressure to project the velocity back onto the divergence-free manifold, enforcing the incompressibility condition (see App.~\ref{app:time evolution} for details). Then, we solve for the velocity at the next time instant by integrating the momentum equation [Eq.~\eqref{eq:NS_eq_a}] without its pressure term. 
This iterative splitting procedure is equivalent to directly solving Eqs.~\eqref{eq:NS} (see App.~\ref{app:time evolution}).
The resulting velocity field satisfies neither the no-slip nor the incompressibility conditions, but this is corrected for in (steps 2 and 3 of) the next iteration (and, after the last iteration, we run one extra round of steps 2 and 3).
The fifth and final step is to retrieve the solution from its MPS encoding, explained in Sec. \ref{subsection:sampleoracle}.

The Poisson equation is the most computationally-expensive task in each iteration (see Sec.~\ref{subsection:Complexity analysis}).
We tackle this with a DMRG-type algorithm that solves an equivalent linear system~\cite{oseledets2012solution}.
As for the time integration, we use the Euler explicit time stepping, with time step chosen according to known numerical-stability criteria (see App.~\ref{app:time evolution}). 
Finally, every operation on an MPS typically increases its bond dimension. We keep it under a chosen threshold by periodic truncation (see App.~\ref{app:bond_dim_truncation}).

\subsection{MPS solution oracle}\label{subsection:sampleoracle}
Our solver outputs the solution in the MPS form, requiring up to exponentially fewer parameters than the dense-vector representation. This is specially relevant for turbulent flows at high Reynolds number, where the smallest cell size must be of the Kolmogorov scale $\mathrm{Re}^{-3/4}$~\cite{kolmogorov1941local,pope_2000}. For fields in 3D this means a mesh size $\mathrm{Re}^{9/4}$. However, this raises the question of how to retrieve the solution without mapping the MPS to the dense vector, which can eliminate potential speed-ups gained at  solving the problem. We propose two methods to extract the main features of the solution directly from the MPS: {\it coarse-grained evaluation} and {\it pixel sampling}.  

\begin{figure*}[t]
    \includegraphics[width=\textwidth]{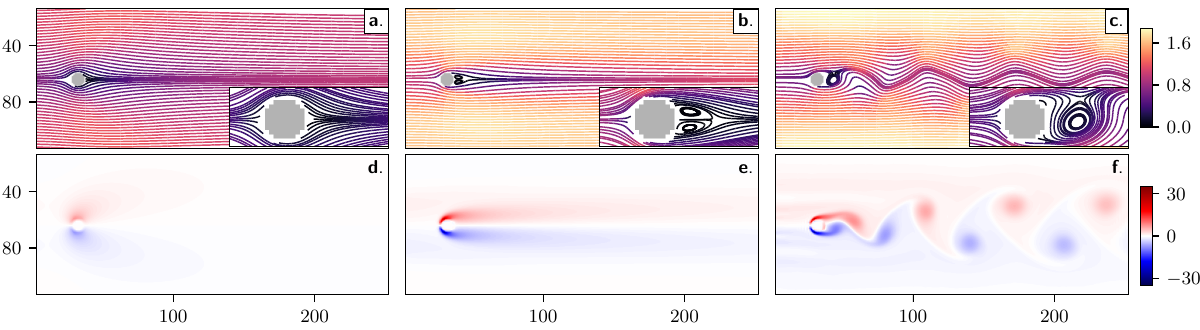}
    \caption[Flow around a circular cylinder at various Reynolds numbers]{\textbf{Flow around a circular cylinder at various Reynolds numbers.} Each column shows one of the three paradigmatic phases of a laminar flow around a cylinder~\cite{Feynman:1494701}.
    Top: streamlines, originating at the inlet (left side), are shown. The color code indicates the magnitude of velocity. The insets show the important features of the flow near the boundary of the object.
    Bottom: vorticity field $\boldsymbol{\omega}=\nabla \times \boldsymbol{v}$ is shown.
    The first column corresponds to $\mathrm{Re}=1.7$: in panel {\sf a}, we observe that the streamlines do not detach from the edge of the cylinder. In turn, in panel {\sf d}, no vorticity is observed behind the object. These are the signatures of a potential flow, where $\boldsymbol{v}$ is given by the gradient of a scalar field.
    The second column corresponds to $\mathrm{Re}=25$. As seen in panel {\sf b}, streamlines detach from the object's boundary. In turn, in panel {\sf e}, a pair of counter-rotating vortices appears behind the object. 
    These are the characteristics of a steady laminar flow.
    The third column corresponds to $\mathrm{Re}=125$. In this regime, the flow becomes unsteady and develops the well-known K\'arm\'an vortex street~\cite{kvs_2004}. This can be observed in panels {\sf c} and {\sf f}, where clear oscillations in the streamlines and in the vortices behind the object are shown.
    A 15-bit MPS with $\rchi = 30$ is used for all cases.
    In all of them, the simulated flow reproduces the expected behaviour.
    }
    \label{fig:cylinder}
\end{figure*}

Coarse-grained evaluation (Fig.~\ref{fig:outline}c) produces an MPS with fewer spatial scales, encoding an averaged version of the solution. 
The averaging is done directly on the MPSs in a highly efficient manner: one simply contracts the physical indices of the fine scales one wishes to average out with single-site constant gadget tensors (in light-green in the figure).
To evaluate the averaged field in full resolution, one then maps the obtained, smaller MPS to its corresponding dense vector contracting the virtual indices between all $N-1$ nearest-neighbours.
Clearly, the finest-scales details are lost in this operation mode. 
A convenient, complementary alternative is pixel sampling.

Pixel sampling   (Fig.~\ref{fig:outline}d) generates random points $(x_{\boldsymbol i},y_{\boldsymbol j})$ chosen with probability $P({\boldsymbol i},{\boldsymbol j})=|v_{\boldsymbol i,\boldsymbol j}|^2/\|{\boldsymbol v}\|_2$, where $\|{\boldsymbol v}\|_2$ is the $l_2$-norm of ${\boldsymbol v}$. 
This reproduces the measurement statistics of a quantum state that encodes ${\boldsymbol v}$ in its amplitudes, hence being potentially relevant also to future quantum solvers~\cite{AsparuGuzik}. We note that related sampling methods have been used for training mesh-free neural-network solvers \cite{SIRIGNANO20181339}. However, there, the points are sampled uniformly, whereas here according to their relevance to the field.  Our approach relies on standard techniques for $l_2$-norm sampling physical-index values from an MPS~\cite{orus2014practical,schollwock2011density}. 
The key ingredient is the marginals of $P({\boldsymbol i},{\boldsymbol j})$, which can be computed by simple single-site tensor contractions on two copies of the MPS (with the gadget tensors in yellow in the figure). 
Remarkably, all $N$ marginals needed are calculated in this fashion without a single evaluation of ${\boldsymbol v}$ at any point $(\boldsymbol i,\boldsymbol j)$. 
With the marginals, the conditional distribution for each bit given the previous ones is obtained, allowing one to sample the entire bit string.
Importantly, the technique applies to any MPS~\cite{ferris2012perfect,han2018unsupervised}. For instance, one can obtain the MPS encoding the vorticity field from that of the velocity. 

\begin{table}[b]
    \centering
    \begin{tabularx}{\columnwidth}{@{}XYY@{}}
         \toprule
         \toprule
         \textbf{Algorithmic task} & \textbf{Complexity} & \textbf{Runtime \%}\\
         \toprule
         Mask generation & $\mathcal{O}(N\,\rchi^2)$ & \text{offline} \\
         \midrule
         Mask application & $\mathcal{O}(\rchi^6)$ & 2.0 \\
         \midrule
         Divergence-free projection & $\mathcal{O}(N\,\rchi^6)$ & \textbf{80.1} \\
         \midrule
         Euler time stepping & $\mathcal{O}(\rchi^6)$ & 17.9 \\
         \midrule
         Coarse-grained evaluation & $\mathcal{O}(N\,\rchi^3)$ & \text{offline} \\
         \midrule
         Pixel sampling (per pixel) & $\mathcal{O}(\rchi^3)$ & \text{offline} \\
         \bottomrule
         \toprule
         \multicolumn{3}{c}{\textbf{Numerically observed total runtime $\approx N\,\rchi^{4.1}$}}\\
         \bottomrule
         \bottomrule
    \end{tabularx}
    \caption[Breakdown of complexity for various steps of the framework.]{\textbf{Time complexities.}
    The first column shows the main subroutines. 
    Their corresponding asymptotic worst-case time complexity scaling and percentages of CPU runtime per time iteration observed are shown in the second and third columns, respectively. 
    The numerical runtimes were measured for the flow around a squared cylinder, using a 15-bit MPS encoding with $\rchi=30$.
    We refer by offline to the tasks performed outside the time evolution loop.
    Finally, the last row shows the scaling of the numerical runtime observed for an entire time iteration (see App.~\ref{app:runtime} for more details).
    }
    \label{table:complexity}
\end{table}

In Fig.~\ref{fig:complexobjects} below, we showcase the real-life applicability of coarse-grained evaluation and pixel sampling for the vorticity field around a wing in a mesh of $2^{19}$ cells. Moreover, in App.~\ref{app:mask}, these evaluation modes allow us to benchmark MPS masks generated for a mesh of $2^{30}$ cells. The hardware used for that is a standard Intel Core i7 CPU, where such high-dimensional benchmark would be impossible using dense vectors.
To end up with, both operation modes can also optionally be combined by {\it close-ups}, i.e. full-resolution evaluations within specifically chosen sub-domains.
Close-ups are obtained fixing a number of large-scale bits of $v_{\boldsymbol i,\boldsymbol j}$, so as to center it on the target sub-domain, and mapping the smaller MPS on the remaining bits to its corresponding dense vector.

\begin{figure*}[t!]
    \centering
    \includegraphics[width=\textwidth]{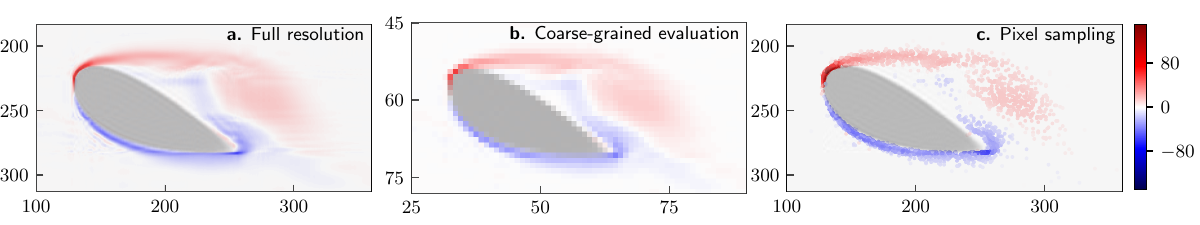}
    \caption[Flow around complex objects]{
    \textbf{Vorticity field around a NACA 0040 airfoil, for different evaluation modes of the MPS solution oracle.} The mesh size is $2^{19}=524288$, the evolution time $T=5000$, and Re $=750$. We compress the corresponding fields into 19-bit MPSs of 18800 parameters, a 27.9X compression. 
{\sf \textbf{a.}} Vorticity field in full resolution. This requires mapping the MPS to the dense-vector representation. {\sf \textbf{b.}} Coarse-grained evaluation, with the two smallest scales in both $x$ and $y$ directions averaged out. The resulting 15-bit MPS has 16 times fewer parameters than the 19-bit MPS solution. {\sf \textbf{c.}} Pixel sampling, featuring 2500 pixels randomly selected according to their squared vorticity. No evaluation of the vorticity field itself is required for that.
}
    \label{fig:complexobjects}
\end{figure*}

\subsection{Computational complexity analysis} \label{subsection:Complexity analysis}

We perform a full time-complexity analysis of the solver and its components. 
Table~\ref{table:complexity} shows the asymptotic (worst-case) upper bounds as well as the numerically observed runtimes, per subroutine. For the tasks within the time integration loop, the estimates are per time iteration.
The complexity of the mask generation is due to the TT-cross approximation~\cite{OSELEDETS201070TTcross}. Those of the other tasks are dominated by their corresponding tensor-network operations (contractions and truncation of $\rchi$)~\cite{schollwock2011density,ferris2012perfect}. 
As for the numerical estimates, they correspond to an average over 10 time iterations.
Note that the projection onto the divergence-free manifold is the most expensive subroutine, with a worst-case performance $\mathcal{O}(N\rchi^6)$. The latter is dominated by a sub-task involving  $N$ linear systems of size $4\rchi^2\times 4\rchi^2$, which we solve exactly since $\rchi$ is low in our cases. However, this can also be solved approximately via variational approaches (see App.~\ref{app:runtime}), which render the complexity $\mathcal{O}(N\rchi^4)$ and are therefore a convenient alternative for high $\rchi$~\cite{gourianov2022quantum}.
Finally, the last row shows the total numerically observed runtime per time iteration. The scaling there is a conservative estimate from numerical fits to the runtimes obtained for fixed $N$ (between $15$ and $23$) as function of $\rchi$ (varying from $20$ to $50$). Those fits are consistent with scalings $\alpha\,N\,\rchi^{4.1}$, with $10^{-7}\leq\alpha\leq 10^{-6}$ (see App.~\ref{app:runtime}). This is significantly below the worst-case bound $\mathcal{O}(N\,\rchi^6)$. 

\begin{figure}[b]
\centering
    \includegraphics[width=\columnwidth]{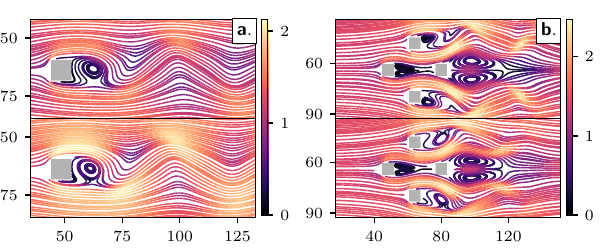}
    \caption[Benchmarking against \texttt{Ansys Fluent}]{\textbf{Comparison with \texttt{Ansys Fluent}.} 
    Panels {\bf a} show the flows around a square cross section for $\mathrm{Re}=127$ and at $T=19900$, obtained by our solver (top) and by the commercial solver \texttt{Ansys Fluent} (bottom). 
    Panels {\bf b} show a similar comparison but for a collection of four squares, for $\mathrm{Re}=141$ and at T=2500. 
    In both panels we take $N_x=8$ and $N_y=7$ as well as $\rchi = 30$ for the MPSs. Importantly, the values of $T$ were chosen high enough for the flows to be in the fully-developed regime (see main text). In both cases the agreement between our solver and \texttt{Ansys Fluent}  is excellent.}
    \label{fig:ansys comparison}
\end{figure}
 
\section{Numerical Results} \label{section:Numerical results}
To showcase the versatility of our framework with three classes of immersed objects. First, we show how our simulations reproduce well-known regimes of flows around a cylinder at different Reynolds numbers. Second, we benchmark our results for flows around squared cylinders against those of the popular commercial solver \texttt{Ansys Fluent}. Third, we tackle the challenging case of flows around an realistic airfoil, where we also showcase the different operation modes of the MPS solution oracle.

The flow around a circular cylinder is one of the paradigmatic problems in fluid dynamics~\cite{kvs_2004}. We simulate the three characteristic dynamics of laminar flow at different $\mathrm{Re}\leq 150$ (see Fig.~\ref{fig:cylinder}). We solve for the velocity and pressure fields on a mesh with $N_x=8$ and $N_y=7$, encoded in a 15-bit MPS with $\rchi=30$ (see App.~\ref{app:bond_dim_truncation} for details). The solutions obtained reproduce the textbook sub-regimes as Re increases: from a flow without downstream circulation to the well-known K\'arm\'an vortex street~\cite{Feynman:1494701}.

Next, we simulate flows around a squared cylinder and a collection of four squared cylinders at $\mathrm{Re}=127$ and $\mathrm{Re}=141$, respectively. Both simulations are performed on a mesh with $N_x=8$ and $N_y=7$ using MPSs of $\rchi = 30$.
As shown in Fig.~\ref{fig:ansys comparison}, we compare our solutions with the ones obtained using \texttt{Ansys Fluent}, observing a remarkable agreement. We note that comparison between conceptually different solvers are meaningful only in the fully developed phase of the flow, after the initial transient behaviour. The evolution times used for Fig.~\ref{fig:ansys comparison} were carefully chosen as to guarantee this condition (see App.~\ref{app:High_Re}).
This rules out possible mismatches in the transient regimes due to different numerical approaches. In addition, in App.~\ref{app:High_Re} we also report simulations for the squared cylinder at $\mathrm{Re}=630$ and $\mathrm{Re}=2230$.

Finally, we consider the simulation of flows around a realistic airfoil in a mesh of $N_x =10$ and $N_y = 9$, shown in Fig.~\ref{fig:complexobjects}. The flow inclination angle used is $22^\circ$, $\mathrm{Re}=750$, and $\rchi=45$. We take advantage of this complex geometry to showcase the different evaluation modes of the MPS solution oracle. 
As a specific example, we consider the vorticity field $\boldsymbol{\omega}=\nabla\times\boldsymbol{v}$. We obtain its associated MPS directly from the velocity MPSs and the matrix-product form of $\nabla$. The vorticity MPS is then taken as the solution oracle for the plots in Fig.~\ref{fig:complexobjects}.
We stress that neither coarse-grained evaluation nor pixel sampling require the vector representation of $\boldsymbol{\omega}$, only its MPS. 

The observed versatility for flows around realistic geometries, together with the built-in ability to access the solution directly from the MPS, renders our framework potentially interesting to hydro-dynamic design problems. In particular, pixel sampling is well-suited for Monte-Carlo estimations of objective functions that may be relevant to industrial optimizations.

\section{Conclusions and Outlook} 
\label{section:Conclusion and Outlook}
We have presented a complete, quantum-inspired framework for CFD based on matrix-product states (MPSs).
Our solver supersedes previous works in that it incorporates immersed objects of diverse geometries, with non-trivial boundary conditions, and can retrieve the solution directly from the compressed MPS encoding, i.e. without passing through the expensive dense-vector representation. For a mesh of size $2^N$, both memory and runtime scale linearly in $N$ and polynomially in the bond dimension $\rchi$.
The latter grows with the complexity of the geometry in question. Nevertheless, we showed that our toolbox can handle highly non-trivial geometries using not only moderate $\rchi$ but also remarkably low parameter numbers. For instance, the most challenging flow simulation considered is for a wing airfoil in a mesh of $2^{19}$ pixels.
Our machinery accurately captures the flows with MPSs of $N=19$ bits, $\rchi=45$, and a total of 18800 parameters (27.9X compression). In turn, it can also accurately represent a cylinder in a mesh of $2^{30}$ cells with an MPS of $N=30$, $\rchi=30$, and a total of 24200 parameters: an astonishing compression of more than 44360X.

A core technical contribution to enable such high efficiencies is the approximate MPS masks. In fact, we have provided a versatile mask generation recipe exploiting standard tensor-network tools such as the TT-cross algorithm. 
Another key ingredient is the alternative methods we have proposed to query the solution  directly from its MPS: coarse-grained evaluation and pixel sampling. 
In particular, pixel sampling explores random mesh points distributed according to the square of a chosen property, such as for instance vorticity or pressure (or, through an equation of state, also temperature).  
This mimics the measurement statistics of a quantum state encoding the property in its amplitudes, hence being relevant also to future quantum solvers~\cite{AsparuGuzik}.
Remarkably, it does so by efficiently computing marginals of the target distribution via local tensor contractions directly on the MPS, without evaluating the property itself at any mesh point. Pixel sampling is ideal for Monte-Carlo simulations, relevant to optimizations via stochastic gradient descent, e.g.
This can be interesting to aero- or hydro-dynamic design problems as well as to training neural-network models. 

Our work offers several directions for future research. In particular, the extension to the 3D case will enable probing the framework in truly turbulent regimes.
This will also require in-depth studies of the dependence of the bond dimension with the evolution time. 
We anticipate that further optimization of the tensor-network Ansatz will play a crucial role for that, as well as extensions to non-uniform meshes. Another possibility is the exploration of tensor networks with other natural function bases for turbulent phenomena, such as Fourier~\cite{li2021fourier} or wavelets~\cite{mariefarge_wavelets}. In turn, an interesting opportunity is the application of our framework to other partial differential equations \cite{Greens_function_MPS,truong2023tensor,Lubasch_multigrid_renorm}. This may be combined with extensions to finite elements or finite volumes, which can in principle also be formulated with tensor networks~\cite{kornev2023chemicalmixer}. Moreover, our method may be relevant to mesh-free solvers too. For instance, pixel sampling on small meshes could be explored to speed-up the training of mesh-free neural-network schemes \cite{SIRIGNANO20181339}. 
Finally, a further prospect could be to combine the framework with quantum  sub-routines, in view of hybrid classical-quantum solvers. 
For example, our main computational bottleneck is the Poisson equation, and future quantum linear solvers may offer significant speed-ups for that~\cite{AsparuGuzik,tosta2023randomized,Lubasch_nonlinear_variational}.

All in all, our findings open a playground with potential to build radically more efficient solvers of real-life fluid dynamics problems as well as other high-dimensional partial differential equation systems, with far-reaching research and development implications.

\bibliography{apssamp}

\clearpage

\setcounter{figure}{0}
\renewcommand\thefigure{\thesection\arabic{figure}}
\setcounter{table}{0}
\renewcommand\thetable{\thesection\arabic{table}}

\onecolumngrid
\begin{center}
    \large \textbf{Supplementary Information}
\end{center}
\twocolumngrid

\appendix
\section{Mask generation}\label{app:mask}
In this appendix, we provide details about the generation of approximate MPS masks outlined in Sec.~\ref{subsection: Mask generation}.

\subsection{$l_2$-distance between $m$ and $\theta$}

We analyze here how well the mask function $m(\boldsymbol{x})$, introduced in Eq.~\eqref{eq:mask}, approximates the indicator function $\theta(\boldsymbol{x})$ of a given object.

\begin{theorem}
    Let the domain be a $D$-dimensional cube of side length $l_0$. Then, for any immerse body geometry, the $l_2$-distance $d_2(\theta,m)$ between the indicator function $\theta$ and its smooth approximation $m$, defined in Eq. \eqref{eq:mask}, goes as $\sqrt{l_0^{D-2}/\alpha}$ for $\alpha\to\infty$.
\end{theorem}

\begin{proof}
    For the 1D case of a segment, $q(x) = x^2 - l_0^2$, where $2 l_0$ is the segment length. Then, the analytical squared $l_2$-distance between $\theta(x)$ and $m(x)$ evaluates to
    \begin{align} \label{eq:l2_distance} \nonumber
        \left[d_2(\theta(x),m(x))\right]^2 & = \int_{-\infty}^{+\infty} (\theta(x)-m(x))^2 dx\\
        &= 2 \int_{l_0}^{+\infty} e^{- 2\alpha (q(x) \nonumber
            + | q(x) |)} dx\\ 
        &=2 \ e^{4\alpha l_0^2} \ \int_{L_0}^{+\infty} e^{-4\alpha x^2} dx \nonumber \\ 
        &= e^{4\alpha l_0^2} \ \sqrt{\frac{\pi}{4\alpha}} \ \mathrm{erfc}(\sqrt{4\alpha l_0^2}) \nonumber \\ & \xrightarrow{a \gg 1}  \frac{1}{4\alpha l_0}~.
    \end{align}
Using similar arguments for the $D$-dimensional cube of side length $l_0$, the $l_2$-distance behaves as $d_2~\sim~\sqrt{l_0^{D-2}/\alpha}$.
\end{proof}
This argument can be extended to arbitrary objects, for which the $l_2$-distance in general is given by $C \alpha^{-\frac{1}{2}}$, where $C$ is independent of $\alpha$. One can indeed fill an arbitrary shape with infinitesimal $D$-dimensional cubes, where each of them carries a constant factor in $\alpha$. Then the resulting $d_2$ will be proportional to the $D$-dimensional object volume times $\alpha^{-\frac{1}{2}}$.
We observe that, for every geometry considered in this work, $\alpha = 10^3$ is sufficient to guarantee a good enough approximation.

\begin{figure}[t]
    \centering
    \includegraphics[width=0.9\columnwidth]
    {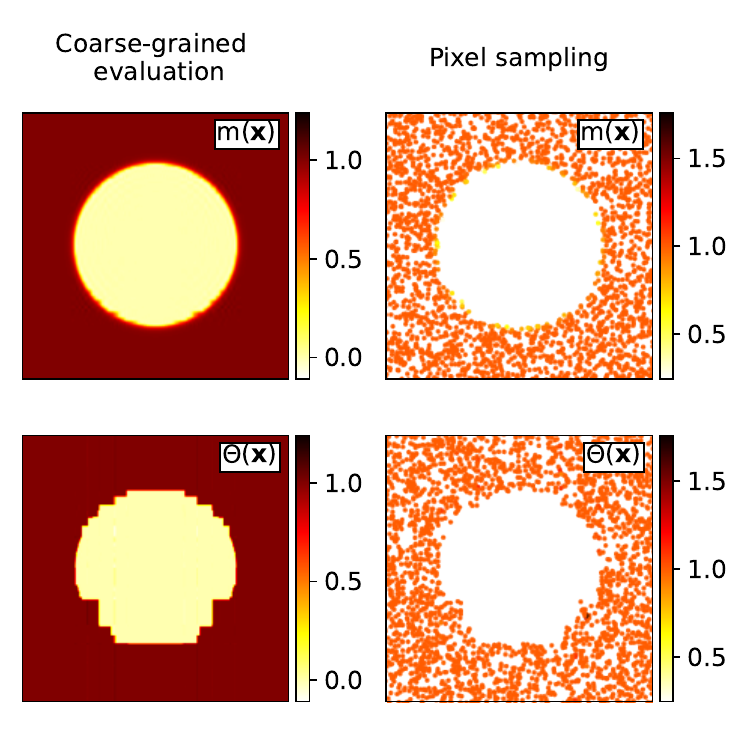}
    \caption[30 bits masks]{\textbf{MPS masks from the indicator function and via its smooth approximation for 30 bits.} Top row: masks generated from smooth approximation $m$. Bottom row: masks generated from indicator function $\theta$.
    Due to the large mesh size, $N_x=N_y=15$, we evaluate the masks using coarse-grained evaluation with 4 scales averaged out (left column) and pixel sampling with $1.2\times10^5$ pixels sampled (right column).
    As can be clearly seen, the masks obtained from $m$ are significantly more accurate than the ones obtained directly from $\theta$.
    In all cases, the bond dimension used is $\rchi = 30$.}
    \label{fig:Masks}
\end{figure}

\subsection{TT-cross approximation to $m$ versus to $\theta$}
We first explain the generation of the MPSs using the TT-cross algorithm~\cite{OSELEDETS201070TTcross}. Then, we present a comparison of MPS masks generated from the indicator function $\theta({\textbf{x}})$ and its smooth approximation $m(\textbf{x})$.

To build an approximate MPS mask for an object, we apply the TT-cross algorithm to the smooth approximation $m$ to the object's boundary's indicator function $\theta$. The TT-cross algorithm builds an MPS approximation to $m$ by querying it on a small number of points $\boldsymbol{x}$.
The queries are made according to a cross-sketching scheme~\cite{OSELEDETS201070TTcross}, without the need to effectively evaluate the dense, exponential-in-$N$, vector representation. 
Then, with the use of acquired queries, TT-cross builds a sequence of CUR decompositions~\cite{bebendorf2000approximation,tyrtyshnikov2000incomplete}.

Fig.~\ref{fig:Masks} shows that the MPS approximation produced by TT-cross is much better when a smoother function like $m(\boldsymbol{x})$ is given as input.
Indeed, in the two columns of Fig.~\ref{fig:Masks}, we plot the two MPS masks arising from the two different application of the TT-cross algorithm for $N_x = 15$, $N_y = 15$ bits domain, using coarse-grained evaluation and pixel sampling respectively.
\begin{figure}[t!]
    \centering
    \includegraphics[width=\columnwidth]
    {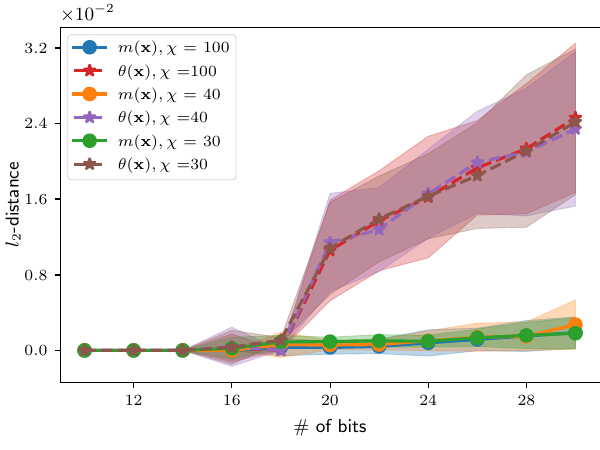}
    \caption[Mask $l_2$ norm]{\textbf{Monte Carlo estimates of $l_2$-distance between the MPS masks and their target functions.}
    The Figure shows that, when using $m(\boldsymbol{x})$ to build the MPS mask, a very good accuracy can be reached even with a constant bond dimension of $\rchi = 30$, independently of the number of bits used. The distance is computed sampling the MPS mask 30000 times and normalizing the $l_2$-distance over the number of samples. 
    Then, since TT-cross is a stochastic process, we run it 30 times and average the results to get a single point in the plot, with its associated standard deviation shown as shaded areas.
    The TT-cross is always run with a limit of $\rchi=100$ in the bond dimension, and the truncation is performed on top of the obtained mask at the end of the process.
    }
    \label{fig:Mask_l2norm}
\end{figure}
\subsection{Bond dimension vs number of bits}
Here, we study the scaling of the bond dimension with the number of bits. In particular, we argue that an MPS mask with a bond dimension $\rchi \leq 30$ is sufficient to capture, i.e. to approximate with a good enough accuracy, the target function $m(\boldsymbol{x})$ for the cylinder case-study considered in this appendix. For the sake of completeness, we conduct the analysis also on MPSs obtained directly from $\theta(\boldsymbol{x})$, for which we find much worse accuracy as expected. In particular, in Fig.~\ref{fig:Mask_l2norm}, we show the $l_2$-distance between the MPS approximation and its target function, which is either $\theta(\boldsymbol{x})$ or $m(\boldsymbol{x})$. Much better approximations are obtained when using $m(\boldsymbol{x})$, confirming the results from the previous section. Moreover, when considering $m(\boldsymbol{x})$, the bond dimension required to achieve a good accuracy does not scale with the number of bits, and we find that $\rchi=30$ is sufficient for our purposes.

\section{Details on the time evolution algorithm} \label{app:time evolution}
\subsection{Differential matrix-product operator with inlet-outlet boundary conditions}
We discuss here the boundary conditions (BC) adopted throughout the work. Since the flow is in the horizontal direction, we apply periodic BCs on the top and bottom edges of the domain. In contrast, on the left and right edges, we apply inlet-outlet BCs~\cite{weinan1995projection}:
\begin{equation} \label{eq:boundarycnds}
    \begin{split}
       \boldsymbol{v}(t)= \boldsymbol{v}_{in}&,~\frac{\partial p(t)}{\partial n}= 0 ,~~\mathrm{at~inlet,}\\
         \frac{\partial v_i(t)}{\partial n} = 0&,~p(t)= p_0,~~\mathrm{at~outlet.}
   \end{split}
\end{equation}
where $\boldsymbol{v}_{in}$ is the inlet velocity, $\partial/\partial n$ denotes partial derivative in the normal direction to the left or right edge of the domain ($\partial/\partial x$ in our case), and $p_0$ fixes the pressure value at the outlet. Eqs.~\eqref{eq:boundarycnds} correspond to a mixed choice of Dirichlet and Neumann boundary conditions, standard in CFD solvers~\cite{anderson1988derivation}.
We employ the ghost cells method~\cite{weinan1995projection} to incorporate the boundary conditions into the matrix representation of the differential operators.
Finally, we map the obtained operator to the MPO form using similar analytical methods to the ones reported in~\cite{kazeev2012low}.

\subsection{Chorin's projection method} 

Here we present the details of the time integration.  We evolve velocity fields within the divergence-free manifold using the Chorin's projection method~\cite{chorin1967numerical,chorin1968numerical} together with finite-difference scheme.%

We begin with the discretization of time in Eqs.~\eqref{eq:NS} stated in main text. 
We choose the explicit Euler time-stepping scheme, resulting in the following:
\begin{equation}
    \label{eq:euler}
    \frac{\boldsymbol{v}_{t+\Delta t} - \boldsymbol{v}_t}{\Delta t}+ (\boldsymbol{v}_t \cdot \nabla )\boldsymbol{v}_t = -\frac{1}{\rho} \nabla p_{t+\Delta t} + \nu \nabla^2 \boldsymbol{v}_t,
\end{equation}
along with the divergence-free condition:
\begin{equation}
    \nabla \cdot \boldsymbol{v}_{t+\Delta t} = 0 .
\end{equation}

Next, we ignore the pressure term in Eq.~\eqref{eq:euler} to produce an intermediate velocity field given by
\begin{equation}
    \label{eq:int_vel}
    \boldsymbol{v}_{t+\Delta t}^* = \boldsymbol{v}_t + (-(\boldsymbol{v}_t \cdot \nabla ) \boldsymbol{v}_t+ \nu\, \nabla^2 \boldsymbol{v}_t)\times\Delta t.
\end{equation}
However, Eq.~\eqref{eq:int_vel} does not satisfy the divergence-free condition.
We then use the Helmholtz decomposition of the intermediate velocity field $\boldsymbol{v}_{t+\Delta t}^*$. This results in two orthogonal components, a solenoidal and an irrotational vector field. By definition, the solenoidal field has zero divergence at all points, which is indeed our objective.

Next, instead of finding the solenoidal component directly, we determine the irrotational component of the intermediate velocity. As stated in the Chorin's projection method~\cite{chorin1967numerical}, this reduces to solving the Poisson equation for the pressure field:
\begin{equation}
    \nabla^2 p_{t+\Delta t} = \frac{\rho}{\Delta t} \nabla \cdot \boldsymbol{v}_{t+\Delta t}^*.
\end{equation}
By subtracting the gradient of the pressure field, we find the solenoidal component of the velocity field, which completes the time evolution for one time step:
\begin{equation}
    \boldsymbol{v}_{t+\Delta t} = \boldsymbol{v}_{t+\Delta t}^* - \frac{\Delta t}{\rho} \nabla p_{t+\Delta t}.
\end{equation}

\subsection{Stability criteria} \label{app:stability criteria}
Along with specifying the domain, mesh size, and boundary conditions, we also need to choose the size of the time step ($\Delta t$) which affects the stability of the time integration. For a stable time evolution, we have to satisfy the Courant–Friedrichs–Lewy (CFL) condition~\cite{cfl_1928}. It states that for ($\frac{U\Delta t}{\Delta x} \leq 1$), where $U$ is the characteristic velocity, the information about the flow travels slower than the flow itself, ensuring stable numerical integration. We emphasize that this is only a necessary condition, but not sufficient for the stability of the algorithm.
However, using this relationship between the temporal resolution ($\Delta t$) and spatial resolution ($\Delta x$), we can estimate an upper bound on the size of the time step.

\section{Truncation of the bond dimension} \label{app:bond_dim_truncation}

\begin{table}[t!]
    \centering
    \begin{tabularx}{\columnwidth}{@{}XYY@{}}
         \toprule
         \toprule
         \textbf{Case study} & $\mathrm{\textbf{Re}}$ & \textbf{Maximum $\rchi$}\\
         \toprule
         Cylinder\\($N_x=8, N_y=7$) & $\mathrm{Re}=125$ & $\rchi = 30$\\
         \midrule
         Squared cylinder\\($N_x=8, N_y=7$) & $\mathrm{Re}=127, 630$ & $\rchi = 30$\\
         \midrule
         Squared cylinder\\($N_x=10, N_y=9$) & $\mathrm{Re}=2230$ & $\rchi = 30$\\
         \midrule
         Many squares\\($N_x=8, N_y=7$) & $\mathrm{Re}=141$ & $\rchi = 30$\\
         \midrule
         Wing\\($N_x=10, N_y=9$) & $\mathrm{Re}=750$ & $\rchi = 45$\\
         \bottomrule
         \bottomrule
    \end{tabularx}
    \caption[Maximum bond dimensions for different simulations.]{\textbf{Maximum bond dimensions for different simulations.} We report all the simulations performed in this work with the associated maximum bond dimension allowed for each case study and the corresponding $\mathrm{Re}$ number.}
    \label{table:maximum_BD}
\end{table}

\begin{figure}[t!]
\centering
 \includegraphics[width=\columnwidth]{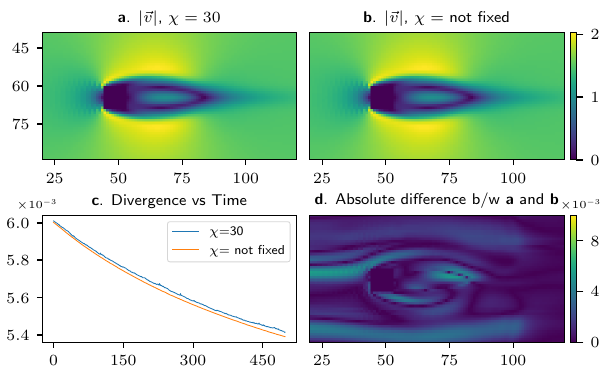}
    \caption[Truncating bond dimension and effects]{\textbf{Comparison of solutions with fixed and unfixed bond dimension $\rchi$.} In panels {\sf a} and {\sf b}, we plot the velocity magnitude of the solutions obtained with a fixed bond dimension of $\rchi=30$ and unfixed $\rchi$ respectively, for the flow around a squared cylinder with $N_x=8$, $N_y=7$ bits. In panel {\sf c}, we show the behaviour of the average divergence ($\nabla \cdot \boldsymbol{v}$) for both cases against time. Panel {\sf d} corresponds to the absolute difference between the velocity fields of panel {\sf a} and {\sf b}.}
    \label{fig:truncation}
\end{figure}

Each time evolution iteration involves several MPS manipulations: element-wise multiplication (mask application and non-linear term in~\eqref{eq:NS_eq_a}), summation (Euler time step), DMRG-type algorithm (Projection step) and MPO-MPS contractions (differential operators). 
Each of them increases the bond dimension of resulting MPSs. Hence, to control the bond dimension, we perform a singular-value truncation of all tensors after each MPS operation~\cite{schollwock2011density}.
In particular, as the first step, we reduce the MPS bond dimension by discarding the smallest singular values such that the $l_2$-norm of the discarded values is below $\epsilon=10^{-16}$. After this, if the bond dimension is still above a certain threshold, we continue to discard the smallest singular values until the threshold is met.
This leads to an increase in the $\epsilon$ value. We remark that we treat this threshold to be a hyper-parameter and tune it such that the increased error during the truncation does not impact the solution. In Table~\ref{table:maximum_BD}, we report the threshold for different case studies.

To illustrate the tuning of this threshold, we present the flow around a squared cylinder. In Fig.~\ref{fig:truncation}, we compare the case with a threshold of $\rchi=30$ (see panel {\sf a}) with the case where only the $\epsilon=10^{-16}$ criterion is used (see panel {\sf b}). In panel {\sf c}, we compare the difference of their corresponding average divergence ($\nabla \cdot \boldsymbol{v}$) as the time evolves. The absolute difference between the velocity fields of panel {\sf a} and {\sf b} is shown in panel {\sf d}. We observe that the MPS solutions resulting from the threshold on bond dimension do not differ significantly from those with no threshold. Hence, we choose the threshold for this case to be $\rchi=30$.

\section{Details about the time complexities} \label{app:runtime}
\subsection{Numerical analysis of runtime}
\begin{figure}[t!]
\centering
 \includegraphics[width=\columnwidth]{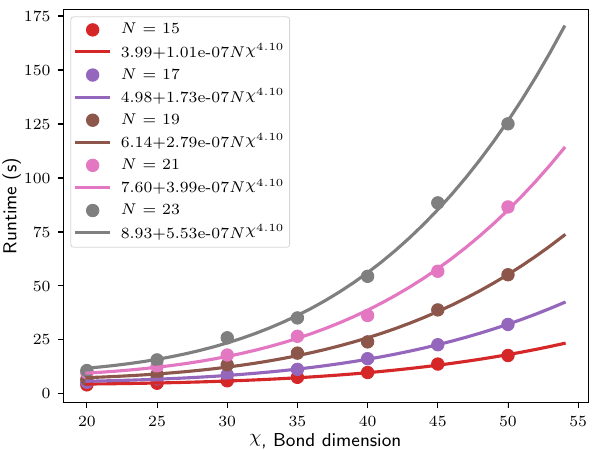}
    \caption{\textbf{Numerical runtimes per one time evolution iteration.} For flow around a squared cylinder, we simulate the evolution using $N$-bit encodings with a bond dimension $\rchi$ and measure the runtime to perform 10 time steps on a standard \texttt{Intel Core i9-10885H} laptop CPU. We plot the average runtime taken per time step against bond dimension for varying number of bits in the encoding. To find the asymptotic behaviour of the runtime, for each MPS encoding, we find the best fit according to the mean squared error constrained on the power being the same across all values of $N$. We report the obtained fits in the legend.}
    \label{fig:runtime}
\end{figure}

Here, we present the details about the estimation of runtime scaling through numerical observations. In Table~\ref{table:complexity}, we present the percentage of runtime taken by each sub-routine as well as the numerically observed scaling. To measure these runtimes, we take the average runtime of 10 time evolution iterations during which the bond dimension does not vary. In Fig.~\ref{fig:runtime}, we fit the average runtimes $t$ (for different fixed values of $N$) as a function of $\rchi$, using the fit Ansatz $t(N,\rchi)=\beta + \alpha\, N \rchi^{\gamma}$. 
Since we are primarily interested in the power-law dependence of $t$ with $\rchi$, we impose the same $\gamma$ for all the values of $N$ we consider, but allow $\beta$ and $\alpha$ to vary for each $N$, for simplicity. Then, we optimize for the mean squared error of the fit simultaneously for all $N$. 
The resulting value of $\gamma$ obtained is $\gamma=4.1$, while values obtained for $\alpha$ range from $10^{-7}$ (for $N=15$) till $5.5 \times 10^{-7}$ (for $N=23$).

\subsection{DMRG-type solver for linear systems}
As identified in Sec.~\ref{subsection:Complexity analysis}, in order to project the velocity fields onto the divergence-free manifold, we solve the resulting linear system using DMRG-based algorithm. This task include two major components: tensor contractions and solution of the ``local'' linear systems. For a fixed bond dimension, the time complexity is then estimated as the combined cost of tensor contractions ($\mathcal{O}(N\rchi^4)$) needed to determine the local linear system of equations of size $4\rchi^2\times 4\rchi^2$, and that of the exact solution of the linear system itself ($\mathcal{O}(N\rchi^6)$), that results in $\mathcal{O}(N\rchi^6)$.

Previous works~\cite{gourianov2022quantum} included variational approaches to tackle the linear system solution with a favorable worst-case complexity $\mathcal{O}(N\rchi^4)$. However, as already emphasised in~\cite{oseledets2012solution}, the exact solution is to be preferred when the bond dimensions involved are small enough (in our cases, $\rchi \le 45$) to allow the direct computation.
In order to reduce the cost of tensor contractions beyond $\mathcal{O}(N\rchi^4)$, we use the \texttt{Random Greedy} optimizer of the \texttt{opt\_einsum} package~\cite{opt_einsum_SG2018}. This finds the contraction order that is approximately minimal in terms of both space and time complexity~\cite{Gray_Kourtis_2021}.

\section{High Re simulations} 
\label{app:High_Re}
\begin{figure}[t!]
\centering
    \includegraphics[width=\columnwidth]{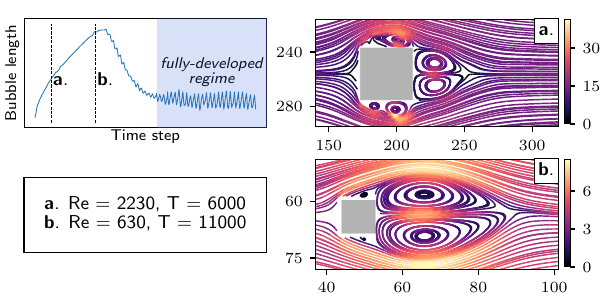}
    \caption[High Re simulations for the squared cylinder] 
    {\textbf{High Re simulations for the squared cylinder.} The top-left panel shows a schematic of the typical behaviour of the bubble length, formed behind the squared cylinder, in time. The shaded region on the right corresponds to the fully developed regime (see panel \textbf{\sf a} in Fig.~\ref{fig:complexobjects}), where oscillations arise. Panel \textbf{\sf a} shows the simulation of the flow around a squared cylinder for $\mathrm{Re}=2230$ on a $N_x=10$, $N_y=9$ mesh with $\rchi=30$. Panel \textbf{\sf b} shows the same case study for $\mathrm{Re}=630$ on a $N_x=8, N_y=7$ mesh with $\rchi=30$. The bottom left panel reports the number of time iterations T performed for both cases.}
    \label{fig:highReflow}
\end{figure}

We report here two simulations for higher Reynolds numbers for the case study of a flow around a squared cylinder introduced in Sec.~\ref{section:Numerical results}.
In particular, we simulate the flow for $\mathrm{Re}=630$ and $\mathrm{Re}=2230$, whose details are reported in Fig.~\ref{fig:highReflow}. We remark that no comparison with \texttt{Ansys Fluent} is presented for these cases, as the flow is still in a developing phase. In fact, no comparison between the two solvers is meaningful until the fully developed phase is reached, due to the different numerical approaches. The different phases of the flow are highlighted in the top left panel of Fig.~\ref{fig:highReflow}.
For these two cases, the simulations are not performed up to the fully developed phase of the flow because we are limited to a small time step $\Delta t$. As explained in App.~\ref{app:stability criteria}, this is due to the CFL condition and the fine mesh dictated by the high Reynolds numbers considered. In order to increase the size of the time step, methods such as implicit time stepping would have to be considered, although at the cost of increasing the runtime per time step. We consider these improvements to be a future direction of work, which will complement the benefits of the proposed framework.

\end{document}